\documentstyle[twocolumn,tighten,aps,epsfig]{revtex}

\begin{document}
\title{Volume 85, Number 4 \hfill PHYSICAL REVIEW LETTERS \hfill 
                                  $\qquad\qquad$ 24 July 2000}
\maketitle
\setcounter{page}{897}


\narrowtext

\noindent{\bf Schulman Replies:}~Casati et al.\ \cite{casati} have failed to understand the goals of my article \cite{opposite}. They and I find ourselves in a situation that is common in discussions of the ``arrow of time," namely no disagreement on technical issues and no agreement on basic assumptions. Instead of criticizing their goals (presumably ``to clear out the too-long standing confusion around (ir)reversibility of statistical laws" \cite{casati}), which occupy many workers in this area \cite{lebowitz}, I will try to make clear what problem I do address.

About 40 years ago Gold \cite{goldajp} argued that the thermodynamic arrow of time followed the cosmological one. This is a thesis of beauty and scope, but I found \cite{correlating} that for logical consistency one needed to introduce a new element into the argument: the use of boundary conditions at two (usually remote) times. This is because the use of macroscopic {\it initial} conditions already fixes the thermodynamic arrow. A further benefit of this conceptual point is that {\it if\/} we are in a big-bang/big-crunch universe and {\it if}, relieved of the initial conditions perspective, one would give roughly symmetric boundary conditions, one would then have a framework within which to look for physical effects arising from the future big crunch. See \cite{timebook}, where these arguments are spelled out in greater detail and the relevant literature, particularly due to J. A. Wheeler, is presented.

A consequence of this framework is that one can imagine subsystems that, because of a particular history of isolation, survive the era of maximal expansion with arrow intact. Such a scenario has been the basis for attacks on Gold's thesis \cite{penrose}. It is this isolation and survival phenomenon that is captured by the boundary conditions given in my article \cite{opposite}. The two times given there are some intermediate times, not particularly near the end points, such that at those times one system is dominated by the conditioning at one end, the other at the other end. If one desired a more complete picture, one could (in principle) arrange boundary conditions near the big bang and big crunch such that relatively isolated chunks of matter would pass the point of maximal expansion with arrow intact. (Obviously this requires greater spatial and dynamical richness than the cat map.) If the time interval under study is $[t_1,t_2]$, then one system would have low entropy at $t_1$, one at $t_2$ (and both times would be significantly distant from the end points). I modeled this by restricting each subsystem to a single coarse grain in its phase space at its respective ``initial" time. But as part of a larger dynamical scheme (i.e., as a slice out of big bang to big crunch boundary value problem), {\it each system would carry a great deal more information encoded in its microscopic coordinates}. This further information ensures that each system possesses its arrow ``before" (by its own clock) its encounter with the other. In terms of the remarks of Casati et al.\ this microscopic information prevents the bi-directional entropy increase envisioned in their criticism.

The goal of many researchers in this area (and one that Casati et al.\ inaccurately ascribe to me in \cite{opposite}) is the elucidation of macroscopic irreversibility in the face of microscopic (near) reversibility. In fact I take for granted that this high purpose can be achieved. My intent is to see whether in our actual world there can be departures from the uniform direction that all (so far) observed systems appear to possess. This requires a formulation, namely two-time conditioning, that does not eliminate the phenomenon by definition (as would happen with initial conditions). Once one has in this way formulated the problem, one can seek actual physical realization. An important aspect of this is the conclusion \cite{opposite} that we could actually see the stuff, contrary to the views of Wiener \cite{wiener}. 

Because a (roughly) time-symmetric universe is the scenario in which opposite-arrow material is most likely to occur (frankly, I can't think of any other reason for it to exist), the ability to identify and characterize such material and the possibility of its observation, have a direct bearing on significant physical questions. 

\smallskip

L. S. Schulman\newline
\indent Physics Department \newline
\indent Clarkson University\newline
\indent Potsdam, New York 13699-5820 \newline
\smallskip
\noindent Received 3 April 2000 \newline

\end{document}